# LATTICE FOR LONGITUDINAL LOW BETA

C. Biscari, LNF-INFN, Frascati, Italy

*Abstract*
The guidelines for a Φ-factory at very high luminosity are described. A preliminary design for a double ring collider, DAΦNE-II, is presented, fulfilling the requirements of luminosity of the order of $10^{34}$ cm$^{-2}$sec$^{-1}$.

## INTRODUCTION

The maximum luminosity at the Φ-energy has been reached in DAΦNE, the Frascati Φ-factory, with values near $10^{32}$ cm$^{-2}$sec$^{-1}$. The super-factory regime, with two order of magnitude higher luminosities, cannot be reached by a simple upgrade of the collider based on stressing the present parameters.

Keeping the DAΦNE main characteristics (double symmetric ring collider in multibunch configuration and flat beams) the design of the super Φ-factory, DAΦNE-II, includes the innovative strong RF focusing principle[1] with enhanced radiation damping and negative momentum compaction. Many contributions to this workshop describe different aspects of the design[2].

This paper presents the general characteristics of a ring whose lattice fulfills all these conditions, and whose layout fits the existing DAΦNE hall (see Fig.1).

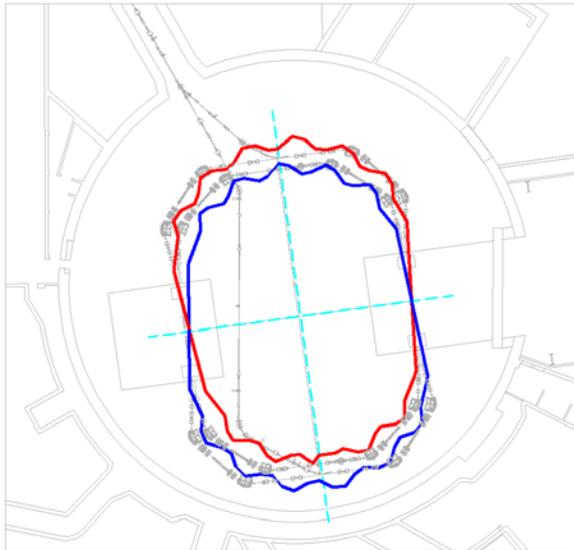

Fig. 1 - DAΦNE-II layout inside the DAΦNE hall

### High radiation damping

Experience in all lepton factories demonstrates the importance of radiation emission and short damping time in obtaining high beam-beam tune shifts. All the low energy rings for lepton collisions are equipped with wigglers to increase the radiation: DAΦNE[3], the LERs of both B-factories[4][5], CESR-c[6], VEPP2M[7].

The damping decrements in terms of the synchrotron radiation integrals $I_n$ (see Appendix) exhibit strong dependence on the energy $E$:

$$\alpha_x = \frac{C_\alpha E^3}{C}(I_2 - I_4)$$
$$\alpha_y = \frac{C_\alpha E^3}{C} I_2 \qquad (1)$$
$$\alpha_s = \frac{C_\alpha E^3}{C}(2I_2 + I_4)$$

with $C_\alpha$ = 2113.1m$^2$/GeV$^3$/s and $C$ the ring circumference.

Presently at DAΦNE the horizontal damping time is of the order of $10^5$ turns (35msec), with almost equal contributions coming from the dipoles (whose bending radius is $\rho$ = 1.4 m) and from the wigglers ($\rho$ = 0.94 m). Since $I_4$ is negligible with respect to $I_2$, the longitudinal damping time is half the horizontal one.

According to beam-beam simulations[8], beam-beam tune shifts higher than the present limit can be obtained at the Φ-energy only if damping times are lowered to few milliseconds, for the typical ring length of 100 m; with respect to the DAΦNE design values, a factor 2 on the beam-beam tune shift corresponds roughly to a factor 10 on the damping time. For a 100 m long isomagnetic ring and neglecting the contribution of $I_4$, the damping time depends on the total length of dipoles or wigglers, $L_{mag}$:

$$\tau_x(sec) = \frac{\rho^2}{2.8 L_{mag}} \qquad (2)$$

with $\rho$ and $L_{mag}$ expressed in meters.

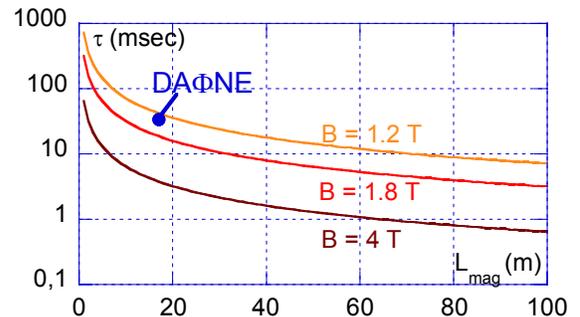

Figure 2: Damping time versus dipole filling factor in a 100 m ring at 510MeV for different magnetic fields. The value of the present damping time for DAΦNE is shown.

This expression is reported in Fig.2 for different values of the magnetic field, including the superconducting option at 4 T: damping times shorter than 10 msec can be obtained only if a large part of the ring is filled with magnets or wigglers, leading to an All-Wiggler-Machine, a concept which was first introduced years ago as a possible solution for synchrotron radiation facilities[9]. A high luminosity collider must be based on a similar solution.

*Negative momentum compaction*

The beam dynamics properties in a ring with negative momentum compaction:

$$\alpha_c = \frac{1}{C} I_1 \qquad (3)$$

are described in a contribution to this workshop[10]. The shorter bunch length and the more regular longitudinal distribution in a lattice with negative $\alpha_c$ are beneficial to the luminosity while the longitudinal beam-beam effect is counteracted by the slope of the RF voltage. All present storage rings work in the positive $\alpha_c$ regime; in some of them, with enough flexibility in the lattice, experiments have been done, setting the machine to a negative momentum compaction configuration. Results confirm simulations: bunch lengths shorter than those corresponding to positive momentum compaction have been obtained[11].

The DAΦNE-II design is based on negative $\alpha_c$.

*Strong RF focusing*

In the flat beam regime the bunch length at the IP must be of the order of the vertical betatron function, i.e. few millimiters for the very high luminosities. Such short bunches cannot be obtained in the low energy rings because of the microwave instability lengthening. The modulation of the bunch length along the ring introduced with the strong RF focusing[1][12] yields very short bunches at the IP while mantaining the average bunch length above the microwave threshold instability, if the impedance sources are concentrated in the high bunch length region. High RF voltage and strong correlation between longitudinal position in the bunch and energy deviation produce the high synchrotron phase advance necessary to focus the beam longitudinally.

## LATTICE CELL

A magnetic lattice has been designed based on the three above principles: high radiation damping, negative momentum compaction, large longitudinal phase advance. The arc structure is based on cells with negative and positive bending magnets[13]. Figure 3 shows the cell layout and fig.4 the betatron functions and the dispersion. In the dipoles the wedge angles correspond to one fourth of the bending angle, and therefore each dipole focuses almost equally in the horizontal and vertical planes.

The dispersion self solution $\eta$ inside the dipoles has the sign opposite to the bending radius' one, so that the contribution to the momentum compaction is negative from all the dipoles. The maximum of the absolute value of the dispersion function occurs inside the magnets and therefore the cell longitudinal phase advance, represented by the term $R_{56}$ of the first order transport matrix, is high.

$$R_{56} = - \int_{dipoles} \frac{\eta(s)}{\rho} ds \qquad (4)$$

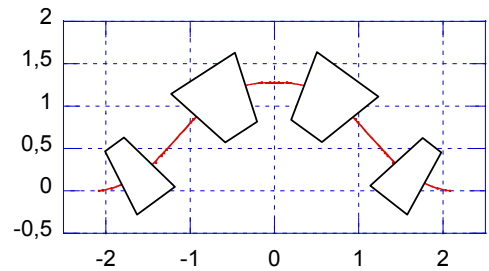

Figure 3 - Schematic cell layout

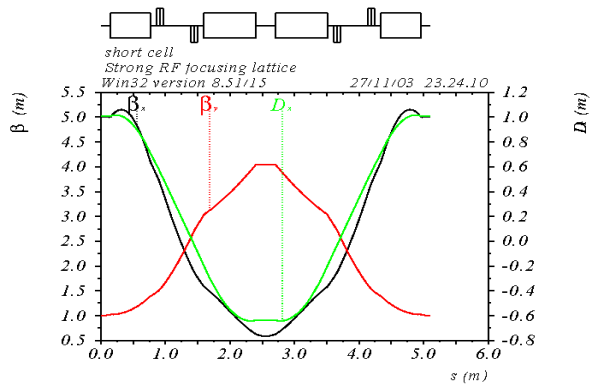

Figure 4 - Cell betatron functions and dispersion

In such a cell the betatron functions are naturally splitted between the dipoles, and chromaticity is corrected by sextupoles placed there, all with the same field orientation, acting alternatively on the vertical plane in the negative dispersion region and on the horizontal plane in the positive one.

The most effective parameter for tuning the momentum compaction is the distance between the positive and the negative dipoles, since it changes the maximum values of the self solution of the dispersion function. By tuning the quadrupole settings, the momentum compaction can still be modified by an amount of the order of 10% without changing significantly the transverse betatron functions. The phase advance per cell is similar in the two planes, and can also be tuned to set the whole ring working point.

# STORAGE RING

The minimum bunch length along the ring occurs at the position where the longitudinal phase advance measured from the rf cavity is half the total one[1]: the $R_{56}$ term between the cavity position and the IP must be the same on both sides of the ring. The ring layout is similar to the DAΦNE one, with a shorter inner arc consisting of five cells and a longer outer of six (see Fig.1). All the cells have the same dipoles and the same bending angle, while the drift lengths and the quadrupole settings are different in the long and short part in order to fulfill the condition:

$$R_{56}^S = \int_{RF}^{IP} \frac{\eta(s)}{\rho} ds \approx R_{56}^L = \int_{IP}^{RF} \frac{\eta(s)}{\rho} ds \quad (5)$$

Dispersion suppressors (DS's) are placed at the arc ends. They are based on one dipole with the same bending radius of the cell dipoles and a quadrupole doublet to match dispersion and $R_{56}$ tunability in the arcs. An example of the DS region between the short arc and the IR is shown in Fig. 5. The other three DS's are similar. Matching between the DS and the IR is performed by means of a quadrupole triplet.

The IR design, described in detail in [14], is based on a 10 m long Interaction Region, with two sets of four symmetric quadrupoles yielding $\beta_x$*=0.5 m, $\beta_y$* between 2 and 4 mm. The Interaction Region is compatible with the present KLOE detector with minor modifications.

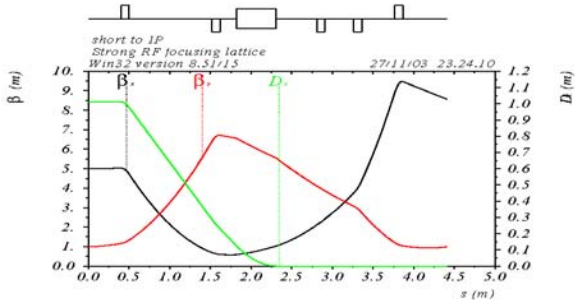

Figure 5 - Betatron functions along the dispersion suppressor near the IR

The long section opposite to the IR has space for housing the RF cavities, the injection septa and kickers[15] and the feedback kickers. Two sets of four quadrupoles match the adjacent dispersion suppressors. In the second crossing point of the rings the beams are vertically separated, and so are the vacuum chambers, in order to avoid any cross-talk between the two beams.

The main parameters of the ring are listed in table I. Figure 6 shows the betatron functions for the whole ring and fig.7 the dispersion function. The Interaction Region is in the central part of the ring while the RF cavity is placed at the beginning.

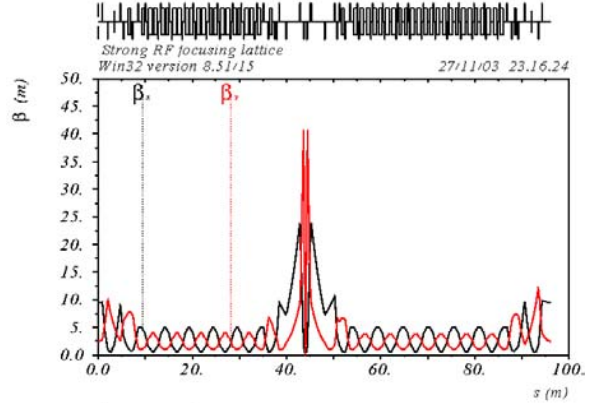

Figure 6 - Betatron functions of the whole ring.

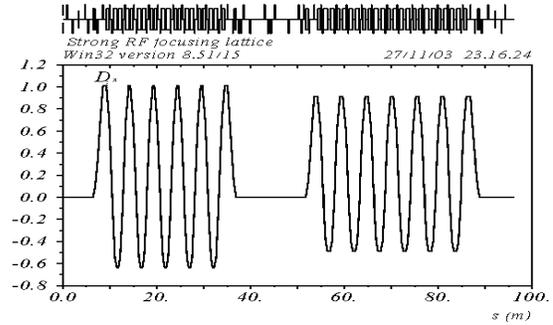

Figure 7 - Dispersion function in m of the whole ring

Table I - DAΦNE-II main parameters

| Parameters | DAΦNE - II |
|---|---|
| $E$ (GeV) | .51 |
| $C$ (m) | 96.1 |
| $L$ ($10^{32}$ cm$^{-2}$s$^{-1}$) | 100 |
| # IPs | 1 |
| $\beta$* (m) (h / v) | 0.5 / 0.003 |
| $\varepsilon$ (μ rad) (h / v) | 0.3 / 0.003 |
| $\theta$ (mrad) | ± 30 |
| $\sigma_z$ (cm) | 0.3-1.4 |
| $\sigma_E/E$ | 1.3 $10^{-3}$ |
| $N_b$ ($10^{10}$) | 5 |
| $\xi$ (h / v) | 0.06 / 0.05 |
| N bunches | 150 |
| I (A) | 3.4 |
| h | 160 |
| $f_{RF}$ (MHz) | 499 |
| V (MV) | 9 |
| $\alpha_c$ | -0.2 |
| $\mu_L$ (°) | 154 |
| # dipoles | 44 |
| B dipoles (T) | 1.8 |
| # quadrupoles | 66 |
| $k_1L$ max (m$^{-1}$) | 1 |
| # sextupoles | 22 |
| $k_2L$ max (m$^{-2}$) | 10 |

# SYNCHROTRON RADIATION AND STRONG RF FOCUSING PARAMETERS

The absolute value of the total bending angle inside one cell is:

$$|\theta_{rad}| = 2 * 48.9° + 2 * 37.5° = 172.72° \quad (6)$$

There are 11 cells in the ring, plus 4 DS; the total dipole length $L_{mag}$ along the ring is:

$$L_{mag} = \rho(11|\theta_{rad}| + \Sigma\theta_{DS}) = 2\pi\rho F_D = 33\ m \quad (7)$$

with $\rho = 0.94$ m[16]. The enhancement factor $F_D = 5.6$ represents the increase of dipole length with respect to a ring with only positive bending radius, at the same magnetic field. The synchrotron radiation integrals are given in the Appendix. The integral $I_4$ is negative and not negligible if compared to $I_2$. It influences strongly the damping partition numbers, and damping is transferred from the longitudinal to the horizontal plane, since:

$$J_x = 1 - \frac{I_4}{I_2} = 1.34$$
$$J_y = 1 \quad (8)$$
$$J_s = 2 + \frac{I_4}{I_2} = 1.66$$

Damping times are therefore: $\tau_x = 7.2$, $\tau_y = 9.6$ and $\tau_s = 5.7$ msec.

Another consequence of the large negative value of $I_4$ is that the natural energy spread is increased: its expression in terms of the synchrotron radiation integrals is given by:

$$\left(\frac{\sigma_E^2}{E^2}\right)_o = C_q \gamma^2 \frac{I_3}{2I_2 + I_4} = (4.9 \cdot 10^{-4})^2 \quad (9)$$

where $C_q = 3.8319 \cdot 10^{-13}$ m. Since the energy spread in the presence of SRFF is higher than the natural one, this is one of the crucial parameters of the design.

The total energy loss per turn is 35 keV and the horizontal emittance is 0.3 mm mrad.

The RF frequency for the ring is near 500 MHz, which is a good compromise between the energy acceptance requirements and the required voltage level[12].

The lattice as described so far can work in a low RF focusing regime, with a voltage of the order of few hundreds of kV and the corresponding bunch length, almost constant along the ring, is of the order of 2 cm for 250 kV with a longitudinal phase advance of ~16°.

By increasing the RF voltage we switch to the strong RF focusing regime. Its complete treatment is described in [1]. The main parameters are here reported. The longitudinal phase advance is given by:

$$\cos\mu = 1 - \pi \frac{\alpha_c C}{\lambda_{RF}} \frac{V_{RF}}{E} \quad (11)$$

The longitudinal beta, is:

$$\beta_l(s) = \frac{\alpha_c C}{\sin\mu_l}\left[1 - 2\pi\frac{R_{56}(s)}{\lambda_{RF}}\left(1 - \frac{R_{56}(s)}{\alpha_c C}\right)\frac{V_{RF}}{E}\right] \quad (12)$$

with $s = 0$ at the RF cavity. $\lambda_{RF}$ the rf wavelength and $V_{RF}$ the rf voltage. The equilibrium energy spread is also related to $\mu$ by:

$$\left(\frac{\sigma_E}{E}\right)^2 = \left(\frac{\sigma_E}{E}\right)_o^2 \frac{2}{1+\cos\mu} \frac{1}{I_3} \oint \frac{\beta(s)_l}{\beta_l(0)|\rho(s)|^3} ds \quad (13)$$

The behaviour of $\beta_l(s)$ along the ring is shown in fig.8 for the longitudinal phase advance $\mu_L = 154°$, corresponding to a RF voltage of 9 MV. The bunch length is shown in fig.9. The minimum bunch length reached in this case is 3 mm and the equilibrium energy spread is $1.3 \cdot 10^{-3}$.

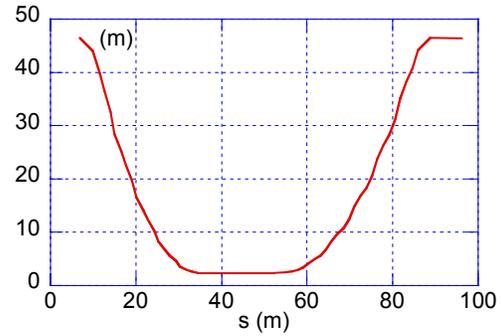

Figure 8 - Longitudinal beta along the ring ($\mu_L = 154°$)

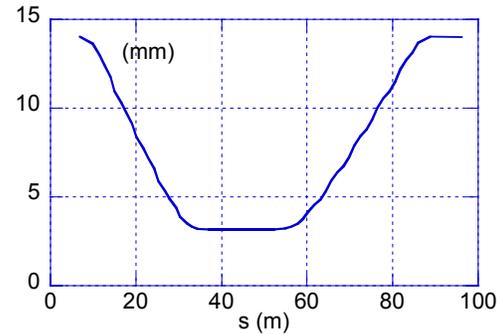

Figure 9 - Bunch length along the ring ($\mu_L = 154°$)

## LUMINOSITY

The single bunch luminosity parameters are listed in table I. The main difference in luminosity with respect to the present DAΦNE values comes from the smaller vertical betatron function, almost by a factor 10, thanks to the bunch length reduction by the same order of magnitude. The bunch current is of the order of 20 mA, similar to the DAΦNE one; further luminosity improvement comes from the smaller emittance and from the lower beam-beam blow-up due to the stronger radiation damping. The higher RF frequency shortens the bunch spacing to 1.8 nsec, and increases the collision frequency. Expected beam-beam tune shifts, neglecting the crossing angle effect, are:

$$\xi_x = \frac{r_e}{2\pi\gamma}\frac{N}{\varepsilon_x} = 0.06 \quad \xi_y = \xi_x\sqrt{\frac{\beta_y}{\beta_x\kappa}} = 0.05 \quad (14)$$

With these parameters and considering that the maximum number of bunches is 150 with an ion clearing gap, the luminosity is expected to reach values near $10^{34}$ cm$^{-2}$sec$^{-1}$.

## DYNAMIC APERTURE

The dynamic aperture of the ring is very large on energy and off-energy without synchrotron oscillations. Sextupoles are naturally placed in the best position for the chromaticity correction. The structure is not critical at low synchrotron tune as well.

On the other hand the preliminary 6-D tracking shows a very strong correlation between the dynamic aperture and the working point in the three phase space planes [17]. In the strong RF focusing regime the dynamic aperture is reduced as compared to the weak focusing case or constant energy deviation. A possible mechanism of this reduction is that the synchrotron motion produces satellites of the strong sextupole resonances and reduces the stable area.

The dependence of the dynamic aperture on the tune point in three dimensions has to be explored.

## CONCLUSIONS

A ring lattice based on the strong RF focusing principle has been designed. The conditions of high radiation damping and negative momentum compaction are also fulfilled.

The ring layout fits the existing DAΦNE hall, and the existing KLOE detector with minor modifications for this one.

A first set of parameters fulfilling the requirement of luminosity in the order of $10^{34}$ cm$^{-2}$sec$^{-1}$ has been defined. The criticality of each parameter must be weighted in order to optimize the overall reliability of the complex.

## APPENDIX

### Synchrotron radiation integrals

The lattice of the DAΦNE-II ring has been calculated with the MAD8 code [18]. The synchrotron radiation integrals are one ot the MAD8 outputs. They have also been evaluated indipendently and a discrepancy has been found in the calculation of the synchrotron integral $I_4$. For rings with achromatic cells and small momentum compaction the value of $I_4$ is negligible with respect to $I_2$ and so are its contributions to the value of emittance and energy spread. This is not the case in the DAΦNE-II lattice. Here we assume that the contribution to $I_4$ by each dipole for an isomagnetic ring is given by the definition of [19]:

$$I_4 = \frac{\tan\theta/4}{\rho^2}(\eta_0 + \eta_1) + \int\frac{\eta(s)}{\rho^3}ds =$$
$$\frac{\tan\theta/4}{\rho^2}(\eta_0 + \eta_1) + \frac{I_1}{\rho^2}$$

where $\theta$ is the dipole angle and $\eta_o$ and $\eta_1$ the dispersion value at the dipole edges. The values of emittance, energy spread and damping partition numbers reported in the paper have been computed with the following values of the synchrotron radiation integrals:

$$I_1 = \oint\frac{\eta}{\rho}ds = -19.2\,m$$

$$I_2 = \oint\frac{ds}{\rho^2} = 37.2\,m^{-1}$$

$$I_3 = \oint\frac{ds}{|\rho|^3} = 39.4\,m^{-2}$$

$$I_4 = -12.2\,m^{-1}$$

$$I_5 = \oint\frac{H}{|\rho|^3}ds = 40.1\,m^{-1}$$